# Limitations in predicting student performance on standardized tests

Michael C. Wittmann, University of Maine, Orono ME 04469-5709, wittmann@umit.maine.edu

The Maryland Physics Expectations Survey (MPEX) describes student attitudes and expectations toward learning, and might be used to predict normalized gains on tests such as the Force and Motion Concept Evaluation (FMCE). These predictions are incomplete, though, due to limitations of the standardized tests themselves. I illustrate the problems involved in using the MPEX to predict productive attitudes toward learning physics by focusing on two students, both with seemingly appropriate expectations toward learning. While one had high normalized gains, the other did not, due to "false favorable" responses on the MPEX.

## Introduction

One way in which physics education research might add value to instruction is by using standardized tests to predict student learning in a course (and tune instruction to an appropriate impedance match). Research-based standardized tests such as the Force and Motion Concept Evaluation (FMCE)[1] could be used to measure the success of our teaching, while tests such as the Maryland Physics Expectations Survey (MPEX)[2] could be used to predict if students have the appropriate attitudes and expectations for successful conceptual learning (as measured by the FMCE). In this paper, I describe a study in which the normalized gains on the FMCE were compared to student responses on the MPEX. The data indicate that, at times, one can predict FMCE gains with the MPEX, while at other times there are "false favorables" in the MPEX which lead to inaccurate predictions of student gains on the FMCE.

## Class Results

The study was carried out at the University of Maine in a sophomore level mechanics course for which differential equations was a corequisite course. Ten students were given the MPEX and then the FMCE on the first day of class and again on the very last day of class. In both cases, they were not aware that they would be tested. During post-instruction testing, students commented that they remembered the tests, but not the individual questions.[3]

Student performance on the FMCE before instruction was weak, and their normalized gains were low (see Figure 1). This is consistent with other courses in which little active learning is used.[1] Since there are no published FMCE data for advanced physics classes, it is hard to compare these data to other similar classes. There are several published results[1] in which higher scores and gains are shown for introductory physics courses in which active learning was used.

Results on the MPEX (see Figure 2) were much more positive. Where students at the end of a typical introductory sequence of physics instruction have scores roughly at (25, 50) for (unfavorable, favorable) responses on the MPEX, these students had better scores initially (pre: 14, 65). Notably, their MPEX results did not deteriorate during the course of the semester. Instead, student responses became less neutral (post: 20, 68), including more favorable and unfavorable responses. Little work has been done with the MPEX in advanced courses to know if these results are typical or anomalous.

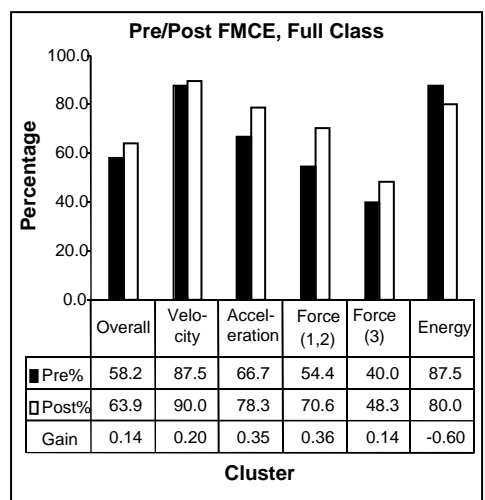

**Figure 1**: Student performance (N=10, matched) on the FMCE before and after all instruction.

I introduce a "normalized movement" plot of MPEX data in figure 2b. The normalization movement is the possible movement to a perfectly favorable score, *e.g.* one in which the responses are all in the upper left corner at (0,100) of the MPEX plot in figure 2a, in line with experts. For the overall score of the class, the possible improvement in scores would be (–14, 35), such that the change in student scores (as shown in figure 2b) is (6/14, 4/35). Figure 2b shows the normalized movement for all clusters on the MPEX,[2] indicating that students gave more favorable and more unfavorable responses in nearly all cases.

## MPEX as an accurate predictor

For some students, the MPEX score can predict normalized gains on the FMCE. Student A was a weak student in the course who had great difficulty connecting the mathematical formalism with the physical descriptions. He began the semester with highly favorable MPEX responses, and his overall score improved slightly from (9,79) to (9, 91) during the semester (see Figure 3). Only on the Coherence cluster did he show no improvement (pre and post: 20, 80). In two clusters (Concepts and Reality link) he began and ended the semester with perfectly favorable scores. In other clusters (Math link and Independence), he ended the semester with perfect scores. Scores, especially in the Concepts, Reality link, and Independence clusters, were consistent with his classroom behavior and my out-of-class interactions with him. Student A's MPEX scores suggest that the student was capable of great improvement in his understanding of the physics.

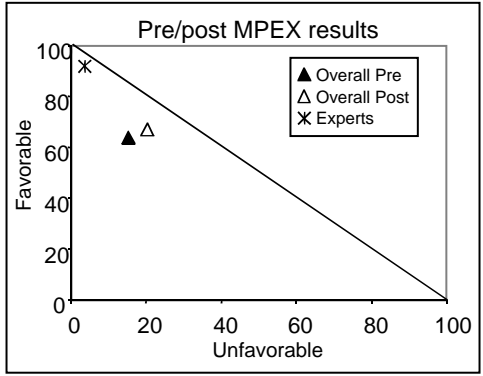

**Figure 2a:** Class performance on the MPEX (overall score) as compared to experts. N = 10 matched students

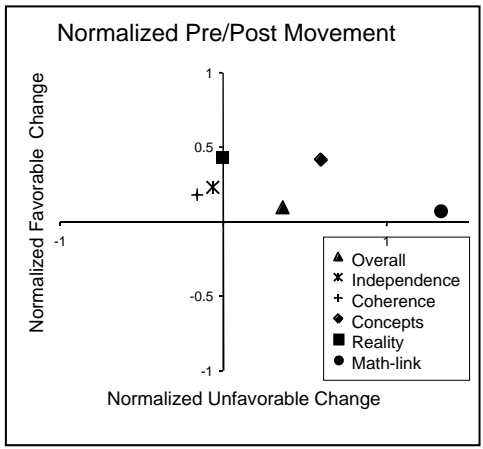

**Figure 2b:** Normalized pre/post MPEX data for 10 matched students. Movement in the second quadrant (favorable scores increase, unfavorable scores decrease) indicates improvement.

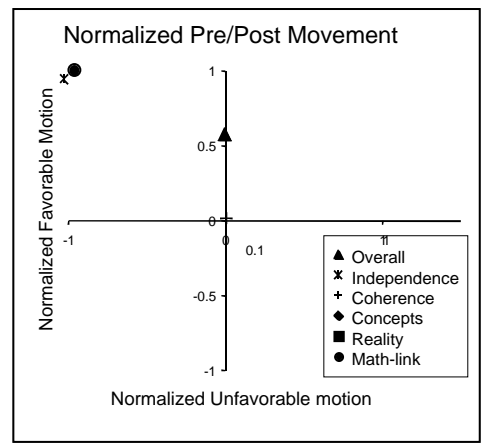

**Figure 3:** Normalized pre/post MPEX data for Student A. Some clusters began and ended perfectly favorable (Concepts, Reality) and are not shown in the graph. Both Math-link and Independence ended with perfectly favorable responses, while Coherence did not change.

This prediction is borne out in his FMCE scores (see Figure 4). Though they began relatively low, they improved greatly in three key areas (acceleration, Newton's $1^{st}$ and $2^{nd}$ laws, and Newton's $3^{rd}$ law). The student's homework and examination scores were consistent with these results. Conceptual understanding of the physics improved greatly; and mathematical understanding lagged behind, but also increased.

## False favorables in the MPEX and inaccurate predictions

Though Student A's results suggest that a favorable MPEX score is an accurate predictor of an FMCE gain, a second example contradicts this conclusion. Student B was weak in physics and mathematics. Her MPEX scores at the beginning of the semester were not unfavorable, though not perfectly favorable. Her overall score changed from (3, 58) to (6, 85) during the semester, indicating a strong improvement in favorable attitudes toward learning physics. She ended the semester with perfectly favorable scores in several clusters (Independence, Reality link, Math link), and showed great improvement in Coherence (0, 40 to 20, 80) and Concepts (0, 40 to 0, 80). The normalized movement of her responses is shown in Figure 5. The general lack of change in unfavorable responses corresponds to the fact that she gave no unfavorable responses on the MPEX before instruction.

Comments made by student B on the MPEX question sheet give reason to question her responses to the Likert-scale questions She often gave favorable responses for unfavorable reasons. For example, she (favorably) disagreed with the question, "All I need to do to understand most of the basic ideas in this course is just read the text, work most of the problems, and/or pay close attention in class," and wrote "Work all of the problems (and do all of that and more)." The response was favorable, but the comment indicated unfavorable attitudes. The comment was consistent with behavior in class and on homework, where she worked all the assigned and additional other end-of-chapter problems by cranking through the math without interpreting

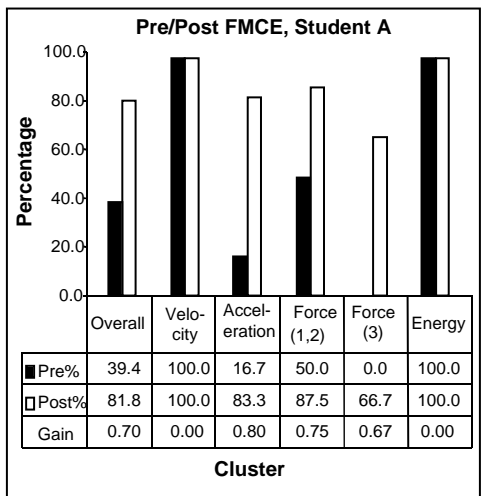

**Figure 4**: Student A pre and post instruction results.

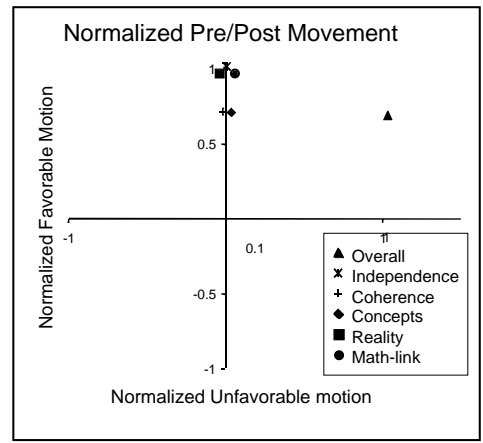

**Figure 5:** Normalized pre/post MPEX data for Student B.

or relating it to the physics.

The student also (favorably) disagreed with the statement "In doing a physics problem, if my calculation gives a result that differs significantly from what I expect, I'd have to trust the calculation." Here, the student wrote, "Very funny. I just go back and try something else." The response was favorable, but the comment reflected strongly unfavorable behavior. In class, the student would see a mistake and start the problem over, without evaluating why it might have gone wrong or checking the steps of the problem.

Student B's performance on the FMCE was the worst in the course. At the start of the semester, only one question was answered correctly, in the velocity cluster of questions. Both at the start and the end of the semester, she had typical incorrect responses in which she, for example, confused the velocity and acceleration of a coin tossed in the air. She improved (notably in the energy cluster, which does not count to the overall score[4]), but had extremely low scores even after instruction.

If one chose to trust the MPEX results (by ignoring the student's comments), one might assume that her pre-instruction knowledge of force and motion was so low that little gain was possible, even in light of favorable attitudes toward learning. But, in light of the student's MPEX comments, it is more likely that the student had unfavorable attitudes toward learning the material. The false favorable responses on the MPEX otherwise lead to an incorrect prediction of student B's possible gains on the FMCE.

## Discussion

In this paper, I hope to have shown that the MPEX is an incomplete predictor of possible gains on tests such as the FMCE. False favorables are most likely only one example of problematic responses to the MPEX. Additional research on the interplay between student attitudes and learning is needed to create a test that avoids false favorables and other possible problems. More importantly, a reliance solely on standardized tests is inadequate when attempting to explore the richness of possible student responses and attitudes toward learning.

## References

[1] Ronald K. Thornton and David R. Sokoloff, "Assessing student learning of Newton's laws: The Force and Motion Conceptual Evaluation and the Evaluation of Active Learning Laboratory and Lecture Curricula," American Journal of Physics **66** (4), 338-352 (1998).

[2] Jeff Saul, Edward F. Redish, and Richard N. Steinberg, "Student expectations in introductory physics," Am. J. Phys. 66, 212-224 (1998).

[3] The scoring was done using analysis templates available at http://perlnet.umephy.maine.edu/materials/. The scoring rubric for the FMCE is based on a personal communication from Ron Thornton.

[4] The energy cluster does not focus on concepts of force and motion, so it is not included in the "overall" score.

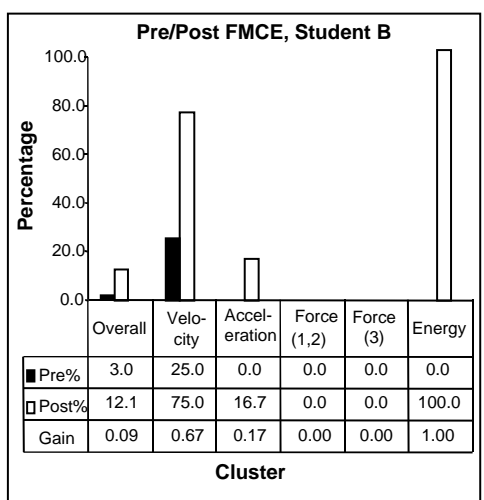

**Figure 6**: Student B pre and post instruction results.

| Cluster | Overall | Velocity | Acceleration | Force (1,2) | Force (3) | Energy |
|---|---|---|---|---|---|---|
| Pre% | 3.0 | 25.0 | 0.0 | 0.0 | 0.0 | 0.0 |
| Post% | 12.1 | 75.0 | 16.7 | 0.0 | 0.0 | 100.0 |
| Gain | 0.09 | 0.67 | 0.17 | 0.00 | 0.00 | 1.00 |